\documentstyle[amssymb,aps,prl,twocolumn,epsfig]{revtex}

\begin{document}

\twocolumn[\hsize\textwidth\columnwidth\hsize\csname
@twocolumnfalse\endcsname

\title{Hole carrier in MgB$_{2}$ characterized by Hall Measurements}
\author{W. N. Kang, C. U. Jung, Kijoon H. P. Kim, Min-Seok Park, S. Y. Lee,
Hyeong-Jin Kim, Eun-Mi Choi, Kyung Hee Kim, Mun-Seog Kim, and Sung-Ik Lee
\cite{email}}
\address{National Creative Research Initiative Center for Superconductivity\\
and Department of Physics, Pohang University of Science and Technology,\\
Pohang 790-784, Republic of Korea }
\draft
\maketitle

\begin{abstract}
The longitudinal resistivity $\rho_{xx}$ and Hall coefficient $R_H$ were
measured for MgB$_2$ sintered under high pressure. We found that $R_H$ is
positive like cuprate high-$T_c$ superconductors, and decreases as
temperature increases for 40 K $< T < $ 300 K. The cotangent of Hall angle
was found to follow $a+b$$T$$^2$ behavior from $T_c$ to 300 K. At $T$ = 100
K, $R_H = 4.1 \times 10^{-11}$ m$^3$/C from which hole carrier density
was determined to be $1.5 \times 10^{23}$/cm$^3$. This carrier
density is 2 - 3 orders of magnitude larger than those of Nb$_3$Sn
and optimally doped YBa$_2$Cu$_3$O$_y$ superconductors.
\end{abstract}

\vskip 0.5pc]

Recently, MgB$_{2}$ was found to be metallic superconductor with transition
temperature ($T_{c}$) of about 40 K.\cite{Akimitsu,CUJung}, and has provided
great scientific interest. Several thermodynamic parameters have been
estimated,\cite{Finnemore,Takano} such as a upper critical field $H_{c2}=$
13 - 18 T, a Ginzburg-Landau parameter $\kappa \sim $ 26, and the critical
supercurrent density $J_{c}(0)\sim $ 10$^{5}$ A/cm$^{2}$. In order to probe
the nature of gap, tunneling spectroscopy measurements have been reported
\cite{Karap,Sharoni}, and they observed superconducting energy gap ($\Delta $%
) of 5 - 7 meV in the framework of the BCS model. The conventional BCS
electron-phonon interaction was proposed as the origin of the
superconductivity based on a band calculation.\cite{Kortus} The possible
origin of the enhanced $T_{c}$ is suggested to originate from a strong
electron-phonon interaction and a enhanced phonon frequency due to the light
boron mass in MgB$_{2}$. Most of the charge carrier density at the Fermi
level comes from the boron band. Indeed, the boron isotope effect has been
reported with an exponent of $\alpha _{B}\sim 0.26$. \cite{Budko} No
experimental study on the electronic structure has been reported yet.

Another interesting feature concerning the normal-state Hall effect in high-$%
T_{c}$ cuprates is the universal temperature dependence of the cotangent of
the Hall angle (cot$\theta _{H}$). Anderson\cite{Chien91} and Chien {\it et
al}.\cite{Anderson91} have proposed that the charge transport is governed by
two different scattering times with different temperature dependences. In
this model, the cot$\theta _{H}$ should be proportional to $T^{2}$ since the
Hall angle is proportional to the inverse of the Hall scattering time $\tau
_{H}$ ($\propto T^{-2}$), and has been observed for most high-$T_{c}$
superconductors.\cite{Ong,Abe} In the mixed-state, the flux-flow Hall effect
is also quite interesting. A puzzling sign anomaly has been observed in some
conventional superconductors\cite{Reed65,Hargen90} as well as in most of
high-$T_{c}$ superconductors.\cite{Kang00} Even double\cite{Budhani,Kang99b}
or triple sign changes\cite{Kang00} have been observed in some high-$T_{c}$
superconductors. Furthermore, a universal scaling behavior between the Hall
resistivity and the longitudinal resistivity has attracted much experimental
\cite{Kang00,Budhani,Kang99b,Kang96} and theoretical interest.\cite{Dorsey92}
However, these Hall effects in the mixed state are not well understood.

To understand the superconductivity in MgB$_{2}$, it is essential to know
the type of charge carrier and it's density, but these have not been
reported yet . Theoretically, Hirsch~\cite{Hirsch} proposed that the 40 K
superconductivity of MgB$_{2}$ originates mainly from the hole carriers with
boron planes acting like the CuO$_{2}$ planes in cuprate high-temperature
superconductors. He proposed that pairing of hole carriers leads to hole
undressing, which is driven by Coulomb interactions. To the best of our
knowledge, the Hall coefficient ($R_{H}$) for MgB$_{2}$ has not been
reported. For other metal diborides (MB$_{2}$, where M = Y, Nb, Ti, V, Cr,
Zr, Mo, Ta, or W), the sign of $R_{H}$ was observed to be negative,\cite
{Hohnson,Juret} and no observation of superconductivity has been reported.
Therefore, the measurement of the Hall effect for MgB$_{2}$ has received
much scientific attention.

To obtain reliable results from transport measurements by using
polycrystalline samples, one must make the sample strong and dense. In this
case, samples sintered under high pressure suitable. In our previous report,
\cite{CUJung} we showed that the mechanical properties, as well as the
superconducting properties, were vastly enhanced for samples sintered at 950
$^{\circ }$C under high pressure (3 GPa range).

In this paper, we report the first measurement of the $R_{H}$ of MgB$_{2}$,
which was carried out using carefully prepared sample with a thin bar shape.
We found that the sign of $R_{H}$ is positive like those of cuprate high-$%
T_{c}$ superconductors. This is contrary to most other metal diborides with
the same structure as MgB$_{2}$. Also the $R_{H}$ decreased as the
temperature increased, and the cotangent of the Hall angle follows $a+bT$$%
^{1.8}$ for most of the measured temperature region from 40 to 300 K. At T =
100 K, $R_{H}=4.1\times 10^{-11}$ m$^{3}$/C, and the calculated hole carrier
density is $1.5\times 10^{23}$/cm$^{3}$.

The polycrystalline samples (4.5 $mm$ in diameter and 3.3 $mm$ in height)
used in this study were sintered at 950 $^{\circ }C$ under 3 GPa. The
fabrication method was reported in detail by Jung {\it et al}.. \cite{CUJung}
The sample purity was more than 99\% as determined by X-ray diffraction
analysis. No grain boundries were observed using the scanning electron
microscopy. In order to obtain a higher Hall voltage signal, we cut the
sample into a bar shape with a length of 4 $mm$ and a width of 2.4 $mm$, and
then mechanically polished it until it was very thin ( 50 - 100 $\mu m$).
The standard photolithography technique was adopted to align the electrical
pads shown in the upper inset of Fig. 1. To obtain good ohmic contacts (%
\mbox{$<$}%
1 $\Omega $), we coated Au film on contact pads after cleaning the sample
surface with an Ar ion beam. This process was done $in-situ$ in a high
vacuum chamber. The voltage noise, which is detrimental to precise
measurements, was successfully reduced to a lower level by preparing a very
thin and optically clean specimen polished from a strong, dense samples.
After installing a low noise preamplifier (N11, EM Electronics ) prior to
the nanovoltmeter (HP 34420A), we achieved a voltage resolution of below 1
nV under a vias current of 50 -100 mA. Fine temperature control was crucial
since the Hall signal was very small. The longitudinal and Hall voltages
were measured simultaneously by using the standard dc 6-probe method. The
magnetic field was applied perpendicular to the sample surface by using a
superconducting magnet system (Maglab2000 Oxford Ltd.) and the applied
current density was $\sim 42$ A/cm$^{2}$. \ The Hall voltage was extracted
from the antisymmetric parts of \ the transverse voltages measured under
opposite directions to remove the longitudinal component due to the
misalignment of the Hall voltage pads. The Hall voltage was found to be
linear in both the current and the magnetic field.

Fig. 1 shows the temperature dependence of the longitudinal resistivity, $%
\rho _{xx}$. The low-field magnetization in the zero-field-cooling state for
the original bulk sample is shown in the lower inset of Fig. 1. The
diamagnetism is 100\% to almost $T_{c}$; thus we normalized to the value at
lower temperature. The superconducting transition temperature is 38.4 K with
a narrow transition width of $\sim $0.6 K, as judged from the 10 to 90\%
superconducting transition. The resistivity value of $\rho \sim $70 $\mu
\Omega $cm at 300 K is comparable to that of single crystalline
intermetallic superconductors.\cite{Fisher} As reported in our previous
work, the normal-state $\rho _{xx}$ follows roughly a $T^{2}$ behavior
rather than a $T^{3}$ behavior, for the entire temperature region below room
temperature.\cite{CUJung} No magnetoresistance was observed from $T_{c}$ to
300 K , which is consistent with the previous results by Jung {\it et al}.
\cite{CUJung} and Takano {\it et al}.,\cite{Takano} but different from the
data by Finnemore {\it et al}..\cite{Finnemore}

The temperature dependence of the Hall coefficient is shown in Fig. 2. The
two curves in the inset represent the Hall voltage measured at 100 K for
opposite magnetic fields up to 5 T. The clearly symmetric and linear shape
demonstrates that the signal to noise ratio for our measurement is high. The
Hall coefficient was positive for all temperatures above $T_{c}$. At 100 K, $%
R_{H}=4.1\times 10^{-11}$ m$^{3}$/C, and the hole carrier density was
calculated to be $1.5\times 10^{23}$ /cm$^{3}$. The absolute value of the
hole carrier density is two orders of magnitude larger than that of Nb$_{3}$%
Sn superconductors\cite{Nolscher} and nearly three orders of magnitude
larger than that of optimally doped YBa$_{2}$Cu$_{3}$O$_{y}$.\cite{Ong}

Hirsch offers an explanation based on a universal mechanism by assuming that
superconductivity in MgB$_{2}$ is similar to that in cuprate superconductors
and is driven by pairing of heavily dressed hole carriers in a band that is
almost full, whereby they gain enough kinetic energy to overcome the Coulomb
energy.\cite{Hirsch} Based on this assumption, he claimed that the type of
the charge carrier is positive.

In Fig. 3, we show the temperature dependence of cot$\theta _{H}$ at 5 T. A
good linear fit to $a+bT^{1.8}$, rather than $a+bT^{2}$, is observed for the
temperature range from $T_{c}$ to 300 K, In high-$T_{c}$ cuprates, the
charge transport is governed by two different scattering times with
different temperature dependences\cite{Anderson91,Chien91} According to this
two-scattering-time model, the longitudinal conductivity ($\sigma _{xx})$ is
governed by the transport scattering time $\tau _{tr}$, which is
proportional to $1/T$, whereas the Hall conductivity ($\sigma _{xy}\thicksim
\tau _{H}\tau _{tr}$) follows $1/T^{3}$ since the Hall relaxation rate is
proportional to $1/T^{2}$. As a result, the cot$\theta _{H}$ (= $\sigma
_{xy}/\sigma _{xx}$) should follow an $\thicksim T^{2\text{ }}$law. Our data
is in fair agreement with a cot$\theta _{H}\thicksim T^{2\text{ }}$law,
which is consistent with the observations in most high-T$_{c}$
superconductors.\cite{Abe} However, this observation cannot be explained by
the above model because the data show $\rho _{xx}\thicksim T^{2}$, as shown
in Fig. 1.

In summary, we report the temperature dependence of $R_{H}$ for the recently
discovered binary superconductor MgB$_{2}$ which has a remarkably high
transition temperature. We find that $R_{H}$ is positive like those for
cuprate high-$T_{c}$ superconductors and that the cot$\theta _{H}$ follows $%
a+b$$T$$^{2}$ from $T_{c}$ to 300 K. At T = 100 K, $R_{H}=4.1\times 10^{-11}$
m$^{3}$/C, and the hole carrier density was $1.5\times 10^{23}$/cm$^{2}$. We
discussed the implication of the hole superconductivity based on a recent
model.

\acknowledgments

We appreciate valuable discussion with J. L. Tallon, This work is supported
by the Ministry of Science and Technology of Korea through the Creative
Research Initiative Program.

\begin{figure}[h]
\centering \epsfig{file=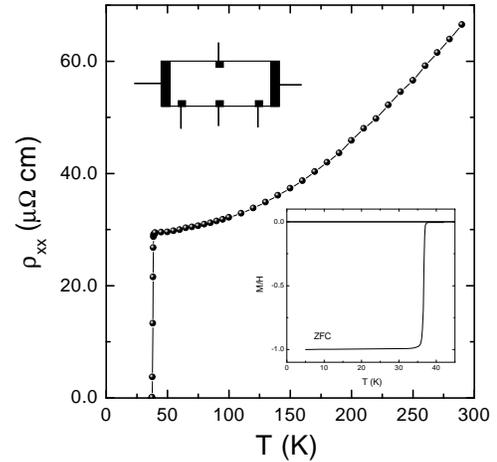, width=6.5cm}
\caption{$\protect\rho _{xx}$ - T shows overall $T^{2}$ behavior and a sharp
transition near $T_{c}$ for MgB$_{2}$ sample. The lower inset shows the
low-field magnetization curve measured in the zero-field-cooling state and
the upper inset shows the configuration of the measurement, namely
six-terminal method. The two electrical pads at both sides are for current
path and the other four leads for longitudinal and transverse voltage
measurement.}
\label{RT}
\end{figure}

\begin{figure}[h]
\centering \epsfig{file=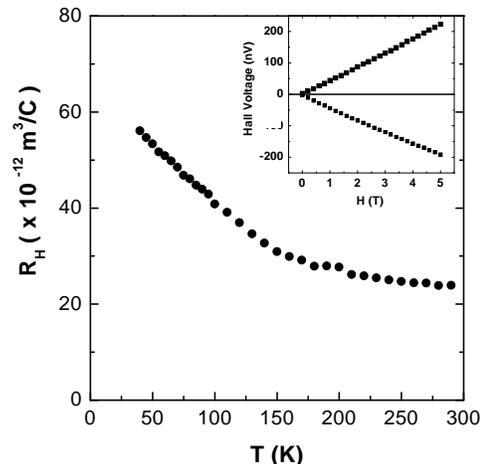, width=6.5cm}
\caption{Hall coefficient measured at 5 T. The two lines in the inset
represent the Hall voltage measured at 100 K for opposite two directions of
the applied field up to 5 T.}
\label{RH}
\end{figure}

\begin{figure}[h]
\centering \epsfig{file=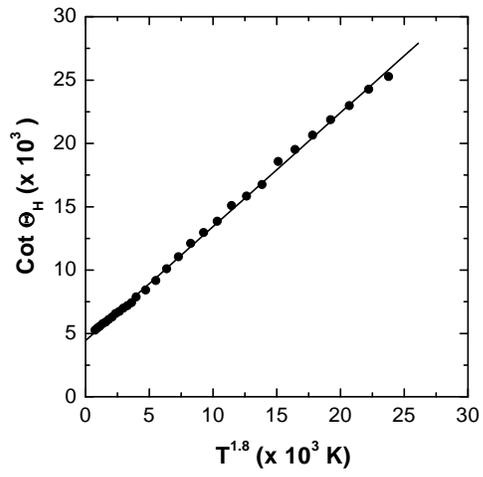, width=6.5cm}
\caption{Cotangent of Hall angle measured at 5 T. The curves show nearly $%
T^{1.8}$ behavior over the entire temperature region measured.}
\label{angle}
\end{figure}

\end{document}